\documentclass[preprint,12pt,authoryear]{elsarticle}

\usepackage{tikz}
\usetikzlibrary{positioning, arrows.meta}
\usepackage[percent]{overpic}
\usepackage[hyphens]{url}

\usepackage{tabularx} 
\newcolumntype{Y}{>{\centering\arraybackslash}X}  
\newcolumntype{Z}{>{\raggedright\arraybackslash}X}  

\usepackage{amssymb}
\usepackage{amsmath}
\usepackage{natbib}
\usepackage{bm}
\usepackage{booktabs}
\usepackage{multirow}
\usepackage{array}
\usepackage{threeparttable}
\usepackage{lineno}
\usepackage{soul}
\usepackage{color}
\usepackage[dvipsnames]{xcolor}

\usepackage[ruled,vlined]{algorithm2e}
\usepackage{makecell}
\sethlcolor{white}

\begin{document}


\begin{frontmatter}



\title{Realistic adversarial scenario generation via human-like pedestrian model for autonomous vehicle control parameter optimisation}

\author[inst1]{Yueyang Wang}
\author[inst2]{Mehmet Dogar}
\author[inst3]{Russell Darling}
\author[inst1]{Gustav Markkula}

\affiliation[inst1]{organization={Institute for Transport Studies},
            addressline={University of Leeds}, 
            city={Leeds},
            postcode={LS2 9JT}, 
            country={UK}}

\affiliation[inst2]{organization={School of Computer Science},
            addressline={University of Leeds}, 
            city={Leeds},
            postcode={LS2 9JT}, 
            country={UK}}
            
\affiliation[inst3]{organization={Five AI Ltd.},
            city={Cambridge},
            postcode={CB1 2JD}, 
            country={UK}}

\begin{abstract}
Autonomous vehicles (AVs) are rapidly advancing and are expected to play a central role in future mobility. Ensuring their safe deployment requires reliable interaction with other road users, not least pedestrians. Direct testing on public roads is costly and unsafe for rare but critical interactions, making simulation a practical alternative. Within simulation-based testing, adversarial scenarios are widely used to probe safety limits, but many prioritise difficulty over realism, producing exaggerated behaviours which may result in AV controllers that are overly conservative. We propose an alternative method, instead using a cognitively inspired pedestrian model featuring both inter-individual and intra-individual variability to generate behaviourally plausible adversarial scenarios. We provide a proof of concept demonstration of this method's potential for AV control optimisation, in closed-loop testing and tuning of an AV controller. Our results show that replacing the rule-based CARLA pedestrian with the human-like model yields more realistic gap acceptance patterns and smoother vehicle decelerations. Unsafe interactions occur only for certain pedestrian individuals and conditions, underscoring the importance of human variability in AV testing. Adversarial scenarios generated by this model can be used to optimise AV control towards safer and more efficient behaviour. Overall, this work illustrates how incorporating human-like road user models into simulation-based adversarial testing can enhance the credibility of AV evaluation and provide a practical basis to behaviourally informed controller optimisation.
\end{abstract}


\end{frontmatter}

\newpage
\section{Introduction}
\label{sec:Introduction}
Autonomous vehicles (AVs) have received significant attention from both academia and industry due to their potential to shape the next generation of mobility. They promise benefits such as reduced labour costs and improved safety by mitigating human error, which accounts for the majority of traffic accidents \citep{muralidhar2023accident}. Major industrial actors such as Waymo and Baidu Apollo have already deployed AV pilots in urban environments, highlighting both the maturity of this technology and the urgent need for its safe evaluation \citep{WaymoUrbanPilot,BaiduApolloGo}.

However, several challenges remain before AVs can be safely deployed at scale in diverse real-world environments. A central requirement is reliable operation in mixed-traffic conditions, where AVs must anticipate and respond to the actions of other road users \citep{camara2020pedestrian}. Evaluating AV behaviour in such interactions is therefore essential prior to large-scale deployment.

Testing these behaviours directly on public roads is costly, time consuming, and potentially hazardous, particularly for rare or safety-critical scenarios. It is therefore impractical to rely solely on real-world testing for validation. Simulation-based testing has consequently emerged as a practical and scalable alternative \citep{huang2016autonomous}. Compared with road testing, virtual environments allow controlled and repeatable experiments across a wide range of conditions, including dangerous or rare interactions \citep{joisher2019simulation,von2023flexible}. For these reasons, simulation is now recognised as a fundamental tool in the design and assessment of AV controller \citep{wang2024survey}.

Within simulation environments, the fidelity of the models that represent other road users is particularly critical, since their behaviour directly shapes the realism and rigour of evaluation. Vulnerable road users, and pedestrians in particular, require special attention as they account for approximately 22\% of global road traffic fatalities~\citep{WHO2013}. However, modelling interactions with pedestrians is particularly challenging due to their high variability and inherent uncertainty. Unlike interactions between vehicles, which often follow traffic regulations and exhibit relatively constrained dynamics, pedestrian behaviour is shaped by a wider set of contextual, psychological, environmental, and motor factors \citep{rasouli2019autonomous}. For instance, differences in motor abilities across individuals influence pedestrians’ walking speed and movement initiation during road crossing \citep{wang2025modeling}. Accurately modelling these subtleties in simulation is therefore essential for avoiding overly simplistic assumptions that limit the real-world applicability of AV control algorithms \citep{mirzabagheri2025navigating}.

Despite this need, most existing simulation frameworks rely on relatively simple pedestrian models. These are often rule-based or deterministic, relying on fixed patterns or basic heuristics such as social-force or cellular automata formulations \citep{helbing1995social,okazaki1979study,blue2001cellular}. While computationally efficient, such approaches oversimplify pedestrian behaviour, failing to capture the natural variability and adaptability that characterise real human decision-making \citep{camara2020pedestrian}. As a result, the realism of pedestrian-vehicle interactions in these simulations remains limited, constraining the validity of the evaluation outcomes.

To improve the realism of pedestrian behaviour representation, more advanced pedestrian models have recently been developed. These include cognitively inspired models that capture perceptual uncertainty and decision-making processes \citep{markkula2018models,pekkanen2022variable,tian2025interacting}, game-theoretic models that represent interactive intent inference \citep{dang2025dynamic}, and reinforcement learning (RL) models that learn bounded optimal behaviour under human-like constraints \citep{wang2025modeling,wang2025pedestrian}. The COMMOTIONS framework, in particular, integrates sensory, motor, and cognitive mechanisms to reproduce more naturalistic pedestrian behaviour \citep{markkula2023explaining}. However, despite these advances, such human-like agents have not yet been systematically integrated into simulation-based AV testing. It therefore remains untested whether applying these human-like pedestrian models within AV simulation testing improves behavioural realism, or how their use might influence the safety-critical behaviours of AVs.

Parallel to these efforts on realism, another line of research has investigated adversarial scenario generation. This refers to the deliberate construction of traffic situations designed to challenge AV decision-making, often by pushing interactions towards unsafe or near-miss outcomes. Such approaches have been investigated primarily as a means to expose failure modes in AV control systems. Typical methods employ RL or black-box optimisation to train pedestrian agents that provoke unsafe interactions, such as collisions with vehicles \citep{song2023critical,hanselmann2022king,priisalu2023varied}. For example, the `suicidal pedestrian' framework explicitly trains agents to collide with AVs \citep{yang2023suicidal}, while the adversarial jaywalker model uses a multi-state rule-based controller to generate hazardous crossings, and has been proposed for AV robustness testing \citep{muktadir2022adversarial}. Although effective for robustness testing of the AV control algorithm, these methods typically prioritise scenario difficulty over behavioural plausibility. The resulting agents often exhibit exaggerated or unrealistic behaviours that fail to reflect real pedestrian intentions, thereby undermining realism and limiting their value for informing real-world deployment \citep{dyro2024realistic}.

Beyond their role in robustness testing, adversarial scenarios also have the potential to inform the optimisation of AV control in pedestrian interaction scenarios. Effective interaction requires balancing safety and efficiency: AVs must avoid collisions and maintain sufficient safety margins, while also preventing unnecessary braking and excessive delays. Optimising AV control against overly simple or overly aggressive pedestrian models may lead to control algorithms that fail to generalise to real human interactions, resulting in behaviours that are either unsafe or overly defensive. In practice, however, adversarial scenario generation has so far been employed mainly as a robustness testing tool, relying on non-human-like pedestrian models that produce exaggerated or unrealistic behaviours. This disconnect reveals a methodological gap: human-like pedestrian models and adversarial testing have largely developed in parallel, and their integration for improving both the realism and practical value of AV evaluation remains limited.

Rather than attempting to replicate the full complexity of real urban environments, the present study focuses on illustrating a conceptual point: that integrating human-like pedestrian models can yield behaviourally plausible adversarial interactions more suitable for AV optimisation than existing rule-based or overly aggressive adversarial agents. As a first step in this line of research, we focus on a simplified one-to-one pedestrian-vehicle interaction at an unsignalised zebra crossing. Building on this foundation, we employ the COMMOTIONS pedestrian model, which integrates human-like sensory, motor, and decision-making mechanisms to reproduce more naturalistic crossing behaviours \citep{markkula2023explaining}. Using this model, we first test whether incorporating a human-like pedestrian agent improves the behavioural realism of AV simulation testing by benchmarking it against the rule-based CARLA pedestrian. We then employ the model to generate adversarial yet behaviourally realistic scenarios, creating challenging but plausible interactions that go beyond prior approaches. Finally, we use these realistic adversarial scenarios to optimise AV control, targeting a trade-off between safety and efficiency, and we benchmark the optimisation outcomes against the adversarial jaywalker baseline \citep{muktadir2022adversarial}.

The novelty of this work lies in unifying human-like behavioural realism with adversarial scenario generation and demonstrating its practical value for AV control. Unlike previous studies that either introduced cognitively inspired pedestrian models without embedding them in AV systems, or used adversarial approaches focused solely on robustness testing, the present study offers a proof of concept demonstration of how closed-loop AV testing and optimisation with human-like pedestrian models can be achieved. Such end-to-end evaluations, where pedestrian models and AV controllers are jointly simulated, remain limited in existing AV research. Through this integration, we analyse how two different AV controllers interact with diverse pedestrian individuals, revealing that safety-critical outcomes emerge only under specific pedestrian instances and conditions. This finding highlights the importance of accounting for both inter- and intra-individual variability in pedestrian behaviour when evaluating AV safety and performance, and demonstrates how behaviourally realistic adversarial scenarios can guide the optimisation of AV control parameters.

\begin{figure*}[!t]
      \centering
      \includegraphics[scale=0.5]{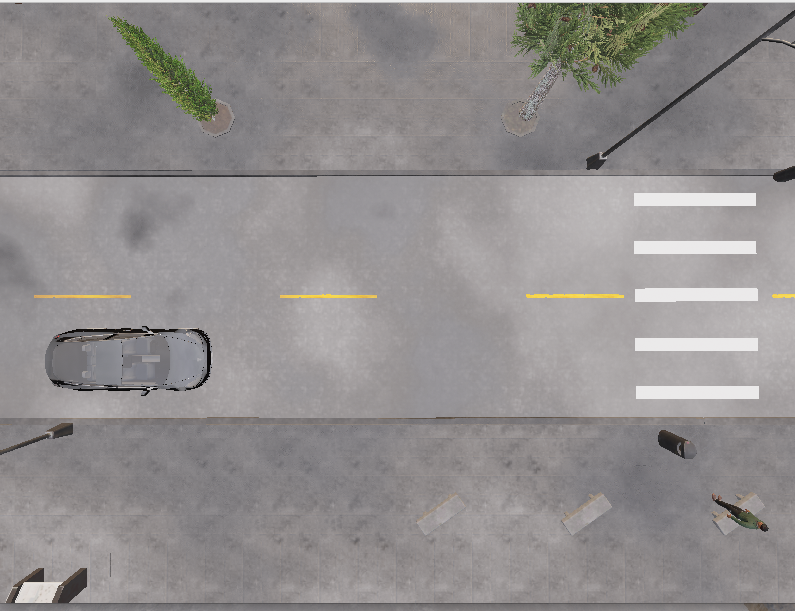}
      \caption{Illustration of the zebra crossing interaction scenario used to evaluate the impact of different pedestrian models on AV behaviour. The pedestrian begins at a standstill, positioned 4 m from the centreline of the vehicle lane and 2 m from the kerb, while the AV approaches at the constant speed under different Time-to-Arrival (TTA) settings.
      The pedestrian began at a standstill, positioned 4\,m from the centreline of the vehicle lane and 2\,m from the kerb, and waited at the kerb before initiating a crossing}
    \label{fig:zebra-crossing-setup}
\end{figure*}

\section{Methods}
\label{sec:Methods}

At the core of our framework is the COMMOTIONS pedestrian model \citep{markkula2023explaining}, which integrates sensory, motor, and cognitive mechanisms to capture human-like road-crossing behaviour. The model was validated by demonstrating that it qualitatively reproduced key interaction phenomena and quantitatively matched pedestrian-driver interaction outcomes observed in a distributed simulator study. It captures inter-individual variability through distinct parameter sets for different individuals, and its perception and decision-making mechanisms are stochastic, i.e., the model also exhibits intra-individual variability. These properties make it suitable both for generating adversarial scenarios that emerge from plausible human variability and for evaluating how AV controllers respond to more realistic pedestrian behaviour. For more detailed information about the model, readers are referred to \cite{markkula2023explaining}. Building on this foundation, our method comprises three main components, which will be described below: (i) evaluation of pedestrian model impact on AV interaction, (ii) generation of adversarial scenarios, and (iii) optimisation of AV control using diverse scenario sets.

\subsection{Evaluation of pedestrian model impact on AV interaction}
\label{subsection:Evaluation of Pedestrian Model Impact on AV Interaction}

The first research question addressed in this study is whether a human-like pedestrian model changes AV-pedestrian interaction outcomes and improves the ecological validity of simulation testing. To support realistic testing of autonomous vehicle behaviour, we integrated the open-source Autoware autonomous driving stack into the CARLA simulation environment via the Autoware-CARLA bridge. This setup allowed the AV to perceive and respond to pedestrians using its full planning and control pipeline, providing a stronger basis for evaluating interactions under different behavioural assumptions.

Using this simulation setup, we compared two pedestrian models and two AV control architectures. The baseline CARLA pedestrian was controlled using the built-in \texttt{WalkerAIController}. As described in the official CARLA documentation \citep{carla_docs_walkers}, the controller places the pedestrian within the simulator’s navigation system, specifies a walking destination, and sets a maximum speed. Once these inputs are provided, the pedestrian’s motion is handled automatically by the simulator’s internal navigation logic, which moves the pedestrian along a feasible path and applies simple reactive avoidance when nearby actors are encountered. The controller does not include decision making, prediction of vehicle motion, or any behavioural model of road crossing.

By contrast, the COMMOTIONS model generates human-like crossing behaviour by integrating perceptual, motor, and cognitive mechanisms that allow adaptive responses to traffic dynamics and variability in behaviour.

In our experiments, for both the CARLA and COMMOTIONS pedestrian models, pedestrians were spawned when the AV reached predefined time-to-arrival (TTA) values, computed from the vehicle’s speed and current distance to the pedestrian crossing point.

Similarly, two AV controllers were considered. The CARLA controller represents a built-in rule-based AV controller that performs lane following and collision avoidance based on predefined thresholds. The Autoware controller, by contrast, uses a full open-source autonomous driving stack, including perception, behaviour planning, and control modules, allowing closed-loop responses to pedestrian motion through sensor-based detection and trajectory planning~\citep{kato2018autoware,autoware_repo}.

We then assessed the influence of pedestrian behavioural realism and AV control architecture using a 2×2 design: pedestrian model (CARLA vs. COMMOTIONS) crossed with AV controller (CARLA planner vs. Autoware).

All experiments were conducted in the CARLA Town~1 map, using a controlled zebra crossing scenario. The pedestrian began at a standstill, positioned 4\,m from the centreline of the vehicle lane and 2\,m from the kerb, and waited at the kerb before initiating a crossing (\figurename~\ref{fig:zebra-crossing-setup}). The autonomous vehicle approached along a straight road at a constant speed of 30 km/h. Vehicles were spawned at a predefined Time-to-Arrival (TTA), defined as the time remaining until the vehicle reached the crossing point. We tested TTAs of 6, 10, 14, and 18 s to systematically vary the level of temporal pressure. By fixing the road layout, vehicle speed, and pedestrian trigger point, the design controlled the temporal and spatial structure of each interaction, ensuring that safety-related outcomes such as collision occurrence and abrupt deceleration events could be evaluated consistently. No other road users were present in the environment, allowing us to focus solely on pedestrian-vehicle interaction.

For the COMMOTIONS pedestrian model, which captures inter-individual variability through distinct parameter sets representing different perceptual/ cognitive/ motor characteristics, five such parameter sets were randomly sampled for each TTA condition. Each individual-TTA combination was then simulated four times to account for intra-individual variability, resulting in 20 simulations per TTA condition. In contrast, the CARLA pedestrian model follows a deterministic behaviour, providing a baseline comparison without inter-individual variation. For comparability, since there is some variability in the AV controller behaviour, the number of runs was nonetheless matched across models: 20 simulations were performed for each TTA condition with both COMMOTIONS and CARLA pedestrians. This ensured that differences in the results reflected behavioural model differences rather than unequal sampling effort.

To evaluate each interaction, we used four behavioural metrics. \emph{Collision rate} captured the proportion of episodes ending in a collision. \emph{Gap acceptance rate} measured how often the pedestrian crossed the road before the car. A crossing was counted as `accepted' if the pedestrian crossed the road before the vehicle reached the conflict point; if the vehicle passed first and the pedestrian crossed afterwards, the trial was counted as `not accepted'. \emph{Post-encroachment time} (PET) is calculated as the time difference between the two agents passing the conflict point \citep{peesapati2018can}. In this study, the conflict point is defined as the geometric intersection of the vehicle’s path along the road and the pedestrian’s crossing path. \emph{Sudden speed change rate} measured how often the vehicle exhibited at least one abrupt deceleration exceeding 2.5 m/s² during a trial, a threshold commonly used in studies of automated driving comfort and safety \citep{carlowitz2024balancing}. These four metrics were chosen to capture complementary aspects of interaction outcomes: safety (collisions, PET), efficiency (gap acceptance), and comfort (abrupt braking).

\subsection{Generation of adversarial scenarios via parameter-TTA search}
\label{subsection:Generation of Adversarial Scenarios via Parameter-TTA Search}

We used the COMMOTIONS pedestrian model together with a parameter-TTA search method to identify individual behavioural profiles and timings that produced critical interactions with low PET. For this experiment, the pedestrian model was paired with the CARLA AV controller rather than the Autoware stack. The CARLA controller can execute large batches of simulations automatically, which is necessary for Bayesian optimisation across hundreds of runs.

For each pedestrian parameter set $p$, the adversarial search problem was formulated as:
\begin{equation}
\min_{\mathrm{TTA}} \quad \mathrm{PET}(p, \mathrm{TTA})
\quad \text{s.t.} \quad \mathrm{TTA} \in [6,18] \ \mathrm{s},
\end{equation}
where $\mathrm{PET}(p,\mathrm{TTA})$ denotes the PET from a simulation with pedestrian parameters $p$ and trigger time $\mathrm{TTA}$, with lower values indicating more critical interactions. 

In practice, we randomly sampled 100 pedestrian parameter sets from COMMOTIONS, each representing a distinct individual with cognitive, motor, and perceptual characteristics. For each individual $p$, Bayesian optimisation was performed independently to identify the TTA value within $[6,18]$\ s that minimised PET. Although in this one-dimensional setting a simple grid search over TTA values would also be feasible, we adopted Bayesian optimisation to use an optimisation procedure that remains effective beyond this specific setup. In higher-dimensional scenario spaces, uniform grid search becomes computationally expensive, whereas Bayesian optimisation is well suited to exploring such spaces efficiently. In addition, PET in our simulations is 
stochastic due to the variability in the COMMOTIONS model, so it is advantageous to use an optimisation method that can handle noisy evaluations.
 
During each optimisation run, a Gaussian-process surrogate model was initialised and updated iteratively with observed PET outcomes. At each iteration, the acquisition function proposed a candidate $\mathrm{TTA}$, the corresponding simulation was executed using the fixed CARLA AV controller and the given pedestrian parameters, and the resulting PET was recorded. The process was repeated until the iteration budget was exhausted, yielding a worst case $\mathrm{TTA}^*$ for that individual. The full procedure is summarised in Algorithm~\ref{alg:ttaopt}. Each optimisation was performed using the \texttt{gp\_minimize} function from \texttt{scikit-optimize}, with 60 iterations and 15 random initial samples.

The resulting set of $\mathrm{TTA}^*$ values—one per individual—defined a set of per-individual minima. Cases with $\mathrm{PET}<1.5$ s were treated as adversarial test sets derived from plausible human behaviour. These adversarial cases formed the basis for the scenario sets used in the subsequent optimisation experiments.

\begin{algorithm}[tb]
\caption{Per-individual Bayesian Optimisation of TTA}
\label{alg:ttaopt}
\KwIn{Set of $N=100$ pedestrian parameter sets $\mathcal{P} = \{p_1,\dots,p_N\}$; 
      TTA range $[6,18]$ s; 
      iteration budget $T=60$ (with $k=15$ random initial samples)}
\KwOut{Optimal TTA values $\{\mathrm{TTA}^*_i\}_{i=1}^N$}
\ForEach{$p_i \in \mathcal{P}$}{
    Randomly sample $k$ initial TTA values from $[6,18]$\;
    \For{$t = 1$ \KwTo $T$}{
        Select next candidate $\mathrm{TTA}_{i,t}$ using Bayesian optimisation\;
        Run simulation with pedestrian parameters $p_i$ and $\mathrm{TTA}_{i,t}$\;
        Compute PET and update the optimisation model\;
    }
    $\mathrm{TTA}^*_i = \arg\min_{\mathrm{TTA}} \mathrm{PET}$ observed for $p_i$\;
}
\Return $\{\mathrm{TTA}^*_i\}_{i=1}^N$
\end{algorithm}

\subsection{Optimisation of AV control using diverse scenario sets}
\label{subsection:Optimisation of AV Control Using Diverse Scenario Sets}

Building on the setup used for adversarial scenario generation, and using the same CARLA AV controller for consistency, we conducted a second experiment to investigate whether risky yet behaviourally plausible scenarios can inform AV control design by focusing the optimisation on a single control parameter: the \textit{braking distance}. This choice was motivated by two considerations. First, the default CARLA AV controller does not implement active yielding behaviour, which limits the options for adapting to pedestrian actions. Second, the aim of this study is to demonstrate a general optimisation framework using behaviourally realistic adversarial scenarios, rather than to fine-tune a complete AV stack. Within this context, braking distance is an appropriate parameter. It defines the threshold distance for initiating an emergency stop: smaller values delay braking and increase risk, whereas larger values trigger earlier braking, improving safety but reducing efficiency. The optimisation task was therefore to identify a value that balances safety and efficiency.

\subsubsection{Scenario sets and rationale}
We optimised the braking distance separately on three complementary scenario sets:

\begin{enumerate}
  \item \textbf{COMMOTIONS low-PET:} The \emph{low-PET subset} from Sec.~\ref{subsection:Generation of Adversarial Scenarios via Parameter-TTA Search}, i.e., the set of individual-TTA pairs that achieved $\mathrm{PET}<1.5$\ s at the per-individual optimum $\mathrm{TTA}^*$. Let its size be $N_H$. These cases correspond to safety-critical interactions arising from human-like pedestrian behaviour.
  
  \item \textbf{COMMOTIONS random:} We drew the same number, $N_H$, of COMMOTIONS individuals and assigned each a TTA sampled independently from a uniform distribution over $[6,18]$ s. This controls for behavioural realism \emph{without} adversarial selection, isolating the added value of targeting low-PET situations.
  \item \textbf{Jaywalker model:} Using the adversarial jaywalker model of \cite{muktadir2022adversarial}, we generated kinematically feasible but not necessarily behaviourally plausible behaviours as a rule-based adversarial baseline. The model cycles through six states (\textit{initialising}, \textit{waiting}, \textit{crossing}, \textit{frozen}, \textit{survival}, \textit{finished}) with social-force dynamics within states, enabling tactics such as sudden dash, mid-road ``freeze'', and ``rewind'' retreat. The jaywalker model was run using the default \textit{risk} mode, which determines its behavioural style in crossing interactions. To make comparisons fair, we matched the number of scenarios to the High-risk set and sampled the jaywalker TTA uniformly from $[6,18]$ s. For further implementation details, readers are referred to \cite{muktadir2022adversarial}.

\end{enumerate}
Together, these sets allow us to compare optimisation driven by (1) high-risk but human-like interactions, (2) realistic yet untargeted behaviour, and (3) the rule-based adversarial jaywalker model.

\subsubsection{AV controller optimisation}

The optimisation problem was designed to balance safety and efficiency. PET was not maximised directly, because very high PET values do not necessarily indicate safer interactions: they may instead reflect that the AV braked excessively early, reducing efficiency or avoiding interaction altogether. Instead, PET was enforced as a constraint with a threshold of 1.5\ s, below which a scenario was deemed unsafe. Likewise, excessive decelerations ($a^{\max}>2.5\ \mathrm{m/s^2}$) were treated as violations of passenger comfort. Within these constraints, the objective was to minimise vehicle time lost, defined as the additional travel time in an interaction run compared to a corresponding free-flow run without the pedestrian.

\begin{algorithm}[tb]
\caption{Bayesian Optimisation of AV Braking Distance}
\label{alg:bayesopt}
\KwIn{Scenario set $\mathcal{S}$; PET threshold $\tau_{\mathrm{PET}}=1.5$\ s; 
deceleration limit $a_{\mathrm{th}}=2.5$\,m/s$^2$; 
search range $d_{\mathrm{brake}} \in [4.0,25.0]$\,m; 
iteration budget $T=60$ with $k=15$ random initial samples}
\KwOut{Optimal braking distance $d^*_{\mathrm{brake}}$}
Sample $k$ initial $d_{\mathrm{brake}}$ values from $[4.0,25.0]$\;
\For{$t = 1$ \KwTo $T$}{
    Select next candidate $d_t$ using Bayesian optimisation\;
    feasible $\leftarrow$ \textbf{true}\;
    Initialise empty list $L$\;
    \ForEach{$(id, \mathrm{TTA}) \in \mathcal{S}$}{
        Run free-flow baseline (no pedestrian) with $d_t$ and measure $T^{\mathrm{free}}$\;
        Run interaction with $d_t$ and pedestrian spawned at TTA; measure $T^{\mathrm{int}}$, compute $\mathrm{PET}$ and $a^{\max}$\;
        \If{$\mathrm{PET} < \tau_{\mathrm{PET}}$ \textbf{or} $a^{\max} > a_{\mathrm{th}}$}{
            feasible $\leftarrow$ \textbf{false}; \textbf{break}\;
        }
        $TL \leftarrow \max(0,\, T^{\mathrm{int}}-T^{\mathrm{free}})$; append $TL$ to list $L$\;
    }
    \eIf{feasible}{objective $\leftarrow$ average($L$)}{objective $\leftarrow$ large\_penalty;}\
    Update optimisation model with $(d_t,\mathrm{objective})$\;
}
\Return $d^*_{\mathrm{brake}} = \arg\min_{d}$ objective;\
\end{algorithm}

The optimisation problem can be expressed as:
\begin{equation}
\begin{aligned}
\min_{d_{\mathrm{brake}}} \quad & \frac{1}{N_{\mathrm{feasible}}} \sum_{i=1}^{N_{\mathrm{feasible}}} T_i \\
\text{s.t.} \quad & \mathrm{PET}_i \geq \tau_{\mathrm{PET}}, \quad \tau_{\mathrm{PET}} = 1.5 \ \mathrm{s}, \\
& a^{\max}_i \leq a_{\mathrm{th}}, \quad a_{\mathrm{th}} = 2.5 \ \mathrm{m/s^2}, \\
& d_{\mathrm{brake}} \in [4.0, 25.0] \ \mathrm{m},
\end{aligned}
\label{eq:optimisation}
\end{equation}
where $T_i$ is the vehicle time lost for scenario $i$, $\mathrm{PET}_i$ is the PET from the interaction run, and $a^{\max}_i$ is the maximum deceleration observed in that run.

The optimisation was run separately for each of the three scenario sets defined above. For every candidate braking distance $d_{\mathrm{brake}}$ proposed by the Bayesian optimiser, the AV was evaluated across all scenarios in the set. Each evaluation involved paired simulations: a free-flow baseline to measure $T^{\mathrm{free}}$, and an interaction run with the pedestrian spawned at the given TTA to measure $T^{\mathrm{int}}$, $\mathrm{PET}$, and $a^{\max}$. A candidate was feasible only if all scenarios met the PET and acceleration thresholds; otherwise, it was penalised with a large objective value. This procedure is summarised in Algorithm~\ref{alg:bayesopt}.

The optimisation employed a Bayesian approach using the Gaussian-process-based \texttt{gp\_minimize} routine from \texttt{scikit-optimize}, with a search space for $d_{\mathrm{brake}}$ defined as [4.0, 25.0]\,m. Each optimisation ran for 60 iterations with 15 random initial samples to balance exploration and exploitation.

\section{Results}
\label{sec:Results}

\subsection{Impact of pedestrian models on AV behaviour}

\begin{figure}[!t]
    \centering
    \includegraphics[scale=0.3]{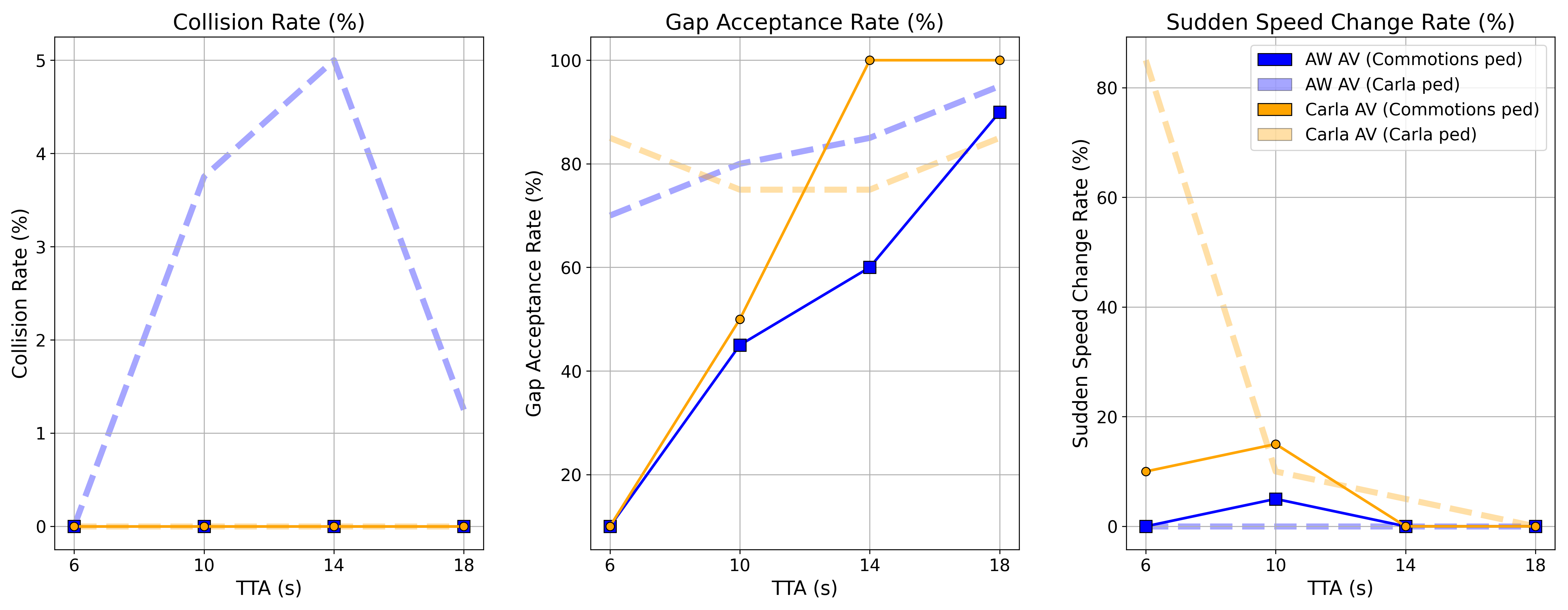}
    \caption{Comparison of AV-pedestrian interaction metrics across different pedestrian and AV model combinations at varying Time-to-Arrival (TTA) values. The three panels from left to right show: collision rate, gap acceptance rate, and sudden speed change rate.}
    \label{fig:metrics}
\end{figure}

From the left panel of \figurename~\ref{fig:metrics}, we can observe that the COMMOTIONS pedestrian model produced no collisions for either the Autoware AV or the CARLA AV across all tested TTA conditions. By contrast, the CARLA pedestrian resulted in multiple collisions, most frequently at intermediate TTAs (10-14 s).

Gap acceptance also differed between the models. For COMMOTIONS, acceptance increased with TTA, starting from about 10\% at 6 s and reaching close to 100\% at 18 s, a progression consistent with findings from earlier studies of pedestrian crossing behaviour \citep{oxley2005crossing,lobjois2007age}. It is worth noting that the TTA values in the present setup are not directly comparable to those reported in the studies where pedestrians make crossing decisions from the kerb, as pedestrians in our simulations start approximately 2\,m from the kerb and therefore require additional time before entering the road. This design choice reflects the fact that real-world interactions often begin before pedestrians reach the kerb \citep{bandini2017collision}. In contrast, the CARLA pedestrian showed no clear trend. When paired with the CARLA AV in particular, the CARLA pedestrian's gap acceptance remained high even at short TTAs such as 6 s, indicating behaviour that does not reflect realistic human sensitivity to the time gap (the middle panel of \figurename~\ref{fig:metrics}).

Sudden speed changes, defined as peak decelerations above $2.5\ \mathrm{m/s^2}$, occurred more often with the CARLA pedestrian testing against Carla AV. At TTA = 6 s, these events exceeded 80\% of trials with the CARLA AV, while with the COMMOTIONS pedestrian they stayed below 20\% across all TTAs and were almost absent with the Autoware AV (the right panel of \figurename~\ref{fig:metrics}).

\begin{figure}[!t]
    \centering
    \includegraphics[scale=0.4]{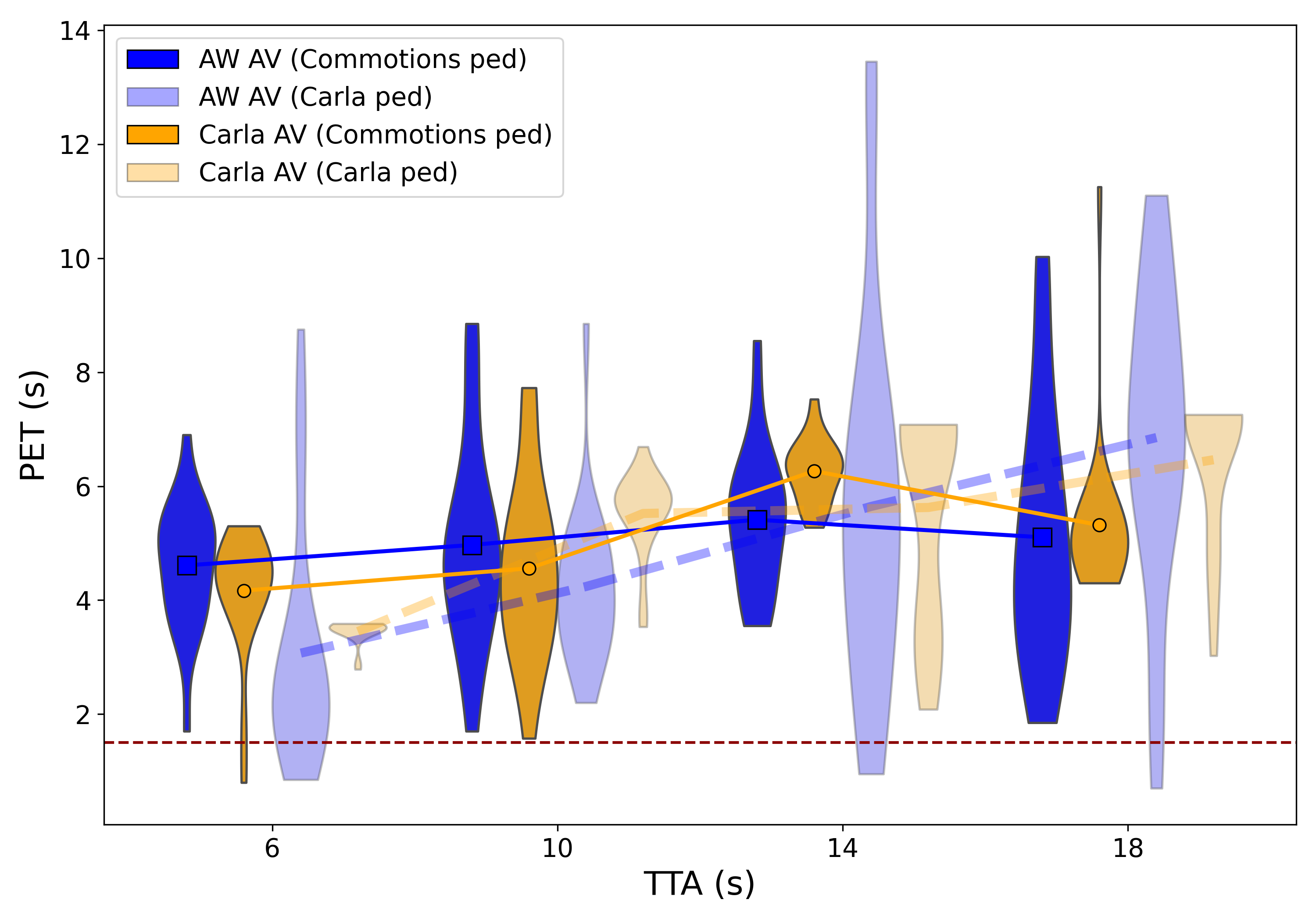}
    \caption{Distribution of post-encroachment times (PET) across pedestrian-AV combinations and TTA values. The red dashed line marks the 1.5 s safety-critical threshold.}
    \label{fig:pet}
\end{figure}

\figurename~\ref{fig:pet} shows the PET across pedestrian-AV combinations and TTA values. For the COMMOTIONS pedestrian, PET values increased with TTA and remained relatively narrow, with almost no samples below the 1.5 s threshold. In contrast, the CARLA pedestrian produced wider variability, most notably at TTA = 14 s, where several PETs dropped below 1.5 s. The mean PET trend lines further illustrate this difference: The COMMOTIONS pedestrian showed a relatively stable increase across TTAs, whereas the CARLA pedestrian exhibited a rise, with much lower mean PET at short TTAs and much higher mean PET at long TTAs.

\begin{figure}[!t]
    \centering
    \includegraphics[scale=0.3]{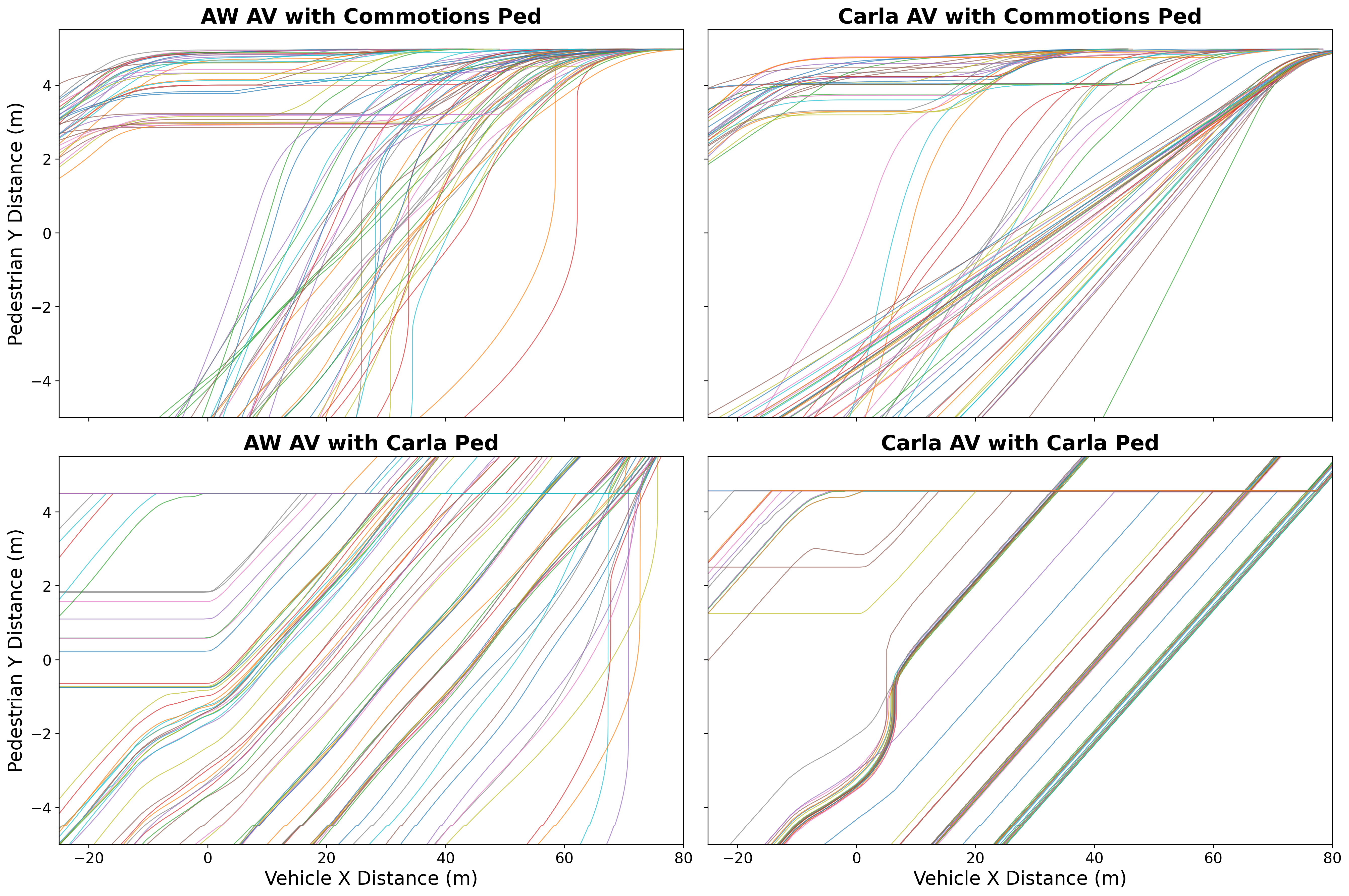}
    \caption{Comparison of pedestrian-vehicle trajectory patterns across different combinations of AV controllers and pedestrian models. Each line represents one interaction, plotted by the vehicle’s and pedestrian’s distances to the crossing point.}
    \label{fig:distance}
\end{figure}

\figurename~\ref{fig:distance} compares the trajectory patterns of pedestrians and vehicles across different combinations of AV controllers and pedestrian models. Each line represents one interaction, with the vehicle’s distance to the crossing point on the x-axis and the pedestrian’s distance on the y-axis. In this plot, each interaction starts from the upper-right corner. The offset in the starting positions along the y-axis for the CARLA pedestrian model is due to implementation constraints: although the pedestrian was intended to be spawned at the same roadside coordinates as in our setup, the CARLA pedestrian can only be placed on the nearest predefined navigation waypoint, which does not always coincide exactly with the specified location. In contrast, the COMMOTIONS pedestrian can be spawned at any specified location.

For both AV controllers, interactions involving COMMOTIONS pedestrians (top panels) show more curved and diverse trajectories, reflecting inter- and intra-individual variability as well as adaptive behaviour and mutual adjustment between the agents. By comparison, trajectories involving the CARLA pedestrian (bottom panels) are mostly straight, indicating limited responsiveness to the approaching vehicle. In addition, the CARLA pedestrian shows more trajectories clustered near the (0, 0) region, suggesting that encounters occurred with shorter temporal and spatial margins. Several horizontal traces around $y = 0$ correspond to the interaction where the CARLA pedestrian became stationary after a collision or a very near miss.

Overall, the above results show that the COMMOTIONS pedestrian generates fewer collisions, smoother AV decelerations, and more consistent PET compared to the CARLA pedestrian, which yields fluctuating gap acceptance and more unsafe outcomes. Together, these results confirm that the COMMOTIONS pedestrian induces richer and more human-like interaction dynamics than the default CARLA pedestrian.

\begin{figure}[!b]
      \centering
      \includegraphics[scale=0.35]{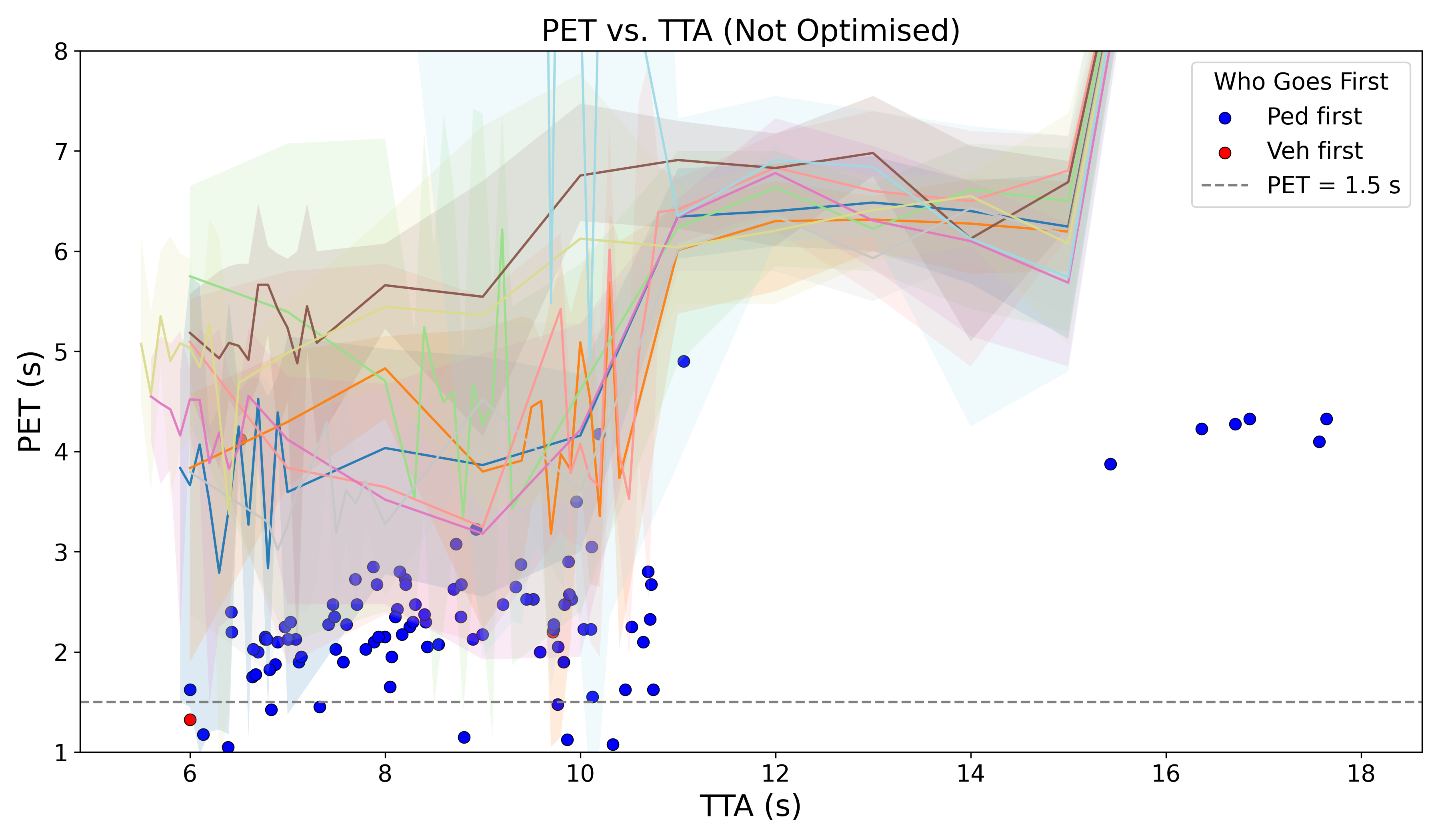}
      \caption{Post-encroachment time (PET) versus Time-to-Arrival (TTA) for the 100 individual pedestrians from the COMMOTIONS model under their respective low-PET individual-TTA combinations prior to optimisation. Markers show the minimum PETs for each individual, and marker colour indicates the order of crossing. Each coloured line shows the mean PET across TTAs for one of the eight pedestrians with PET below the safety threshold of 1.5 s (dashed line), and the shaded envelopes denote the range across repeated runs for these pedestrians, to illustrate stochastic intra-individual variability.}

      \label{fig: low_PET_points}
\end{figure}

\subsection{Identification of safety-critical events}
\label{subsection:Identification of Safety-Critical Events}

As shown in \figurename~\ref{fig:pet}, although the COMMOTIONS pedestrian generally produced safe interactions, an important property of the model is that it doesn't always: a few interactions exhibited PET below the 1.5 s safety threshold. As described in Sec.~\ref{subsection:Generation of Adversarial Scenarios via Parameter-TTA Search}, we used the COMMOTIONS pedestrian model in combination with the CARLA AV controller in a parameter-TTA search procedure to generate low-PET scenarios. The resulting outcomes are shown in \figurename~\ref{fig: low_PET_points}.

\figurename~\ref{fig: low_PET_points} plots PET as a function of TTA for the individuals whose minimum PET fell below 1.5 s. Each line represents one pedestrian individual, with shading indicating the variability across repeated runs arising from variability in the COMMOTIONS model. The plot shows that low PET values typically occurred within the short-TTA range (6-10 s), and it also shows that these low PET values only occurred sometimes for these individuals at these TTAs. In other words, unsafe interactions happen more often for some individuals (inter-individual variability), but not all of the time for any individuals (intra-individual variability).

In \figurename~\ref{fig: low_PET_points}, a clear gap is visible in the 11–15 s TTA range, where individual minima of PET  were observed. For TTAs greater than 15 s, some individuals exhibited their minimal PET values at around 4 s. These individuals are characterised by conservative crossing behaviour and slower walking speeds. Their PET minima occur at large initial TTA values because at these safer TTAs the pedestrians initiate the crossing later and walk more slowly, causing a local minimum in PET.

\subsection{Improvement in the interaction safety and efficiency through AV Parameter optimisation}
\label{subsection:Improvement in the interaction safety and efficiency through AV Parameter Optimisation}

Table~\ref{tab:av-ped-comparison} summarises the results across four conditions: before optimisation, and three post-optimisation approaches based on COMMOTION low-PET, COMMOTION random, and jaywalker model. The values in the table show the results of the different AV controller variants interacting with the same 100 individual-TTA combinations of the COMMOTIONS pedestrian as identified in Sec.~\ref{subsection:Identification of Safety-Critical Events}, ensuring comparability across AV controllers. The rows differ only in the braking distance parameter setting: the default CARLA AV setting for the before-optimisation row, and the optimised values obtained from the three approaches.

When optimisation was guided by low-PET individuals, performance improved across all metrics compared with the before-optimisation baseline. Mean PET almost doubled (4.96 s vs 2.48 s), the frequency of sudden speed changes fell from 83\% to 17\%, and average vehicle time lost was reduced from 0.12 s to 0.05 s. The minimum PET also increased from 1.05 s to 1.80 s, i.e., safety-critical interactions no longer occurred. These results show that tuning on human-like safety-critical interactions enabled the AV to maintain larger temporal margins while braking more smoothly and efficiently.

 \begin{figure}[!b]
      \centering
      \includegraphics[scale=0.5]{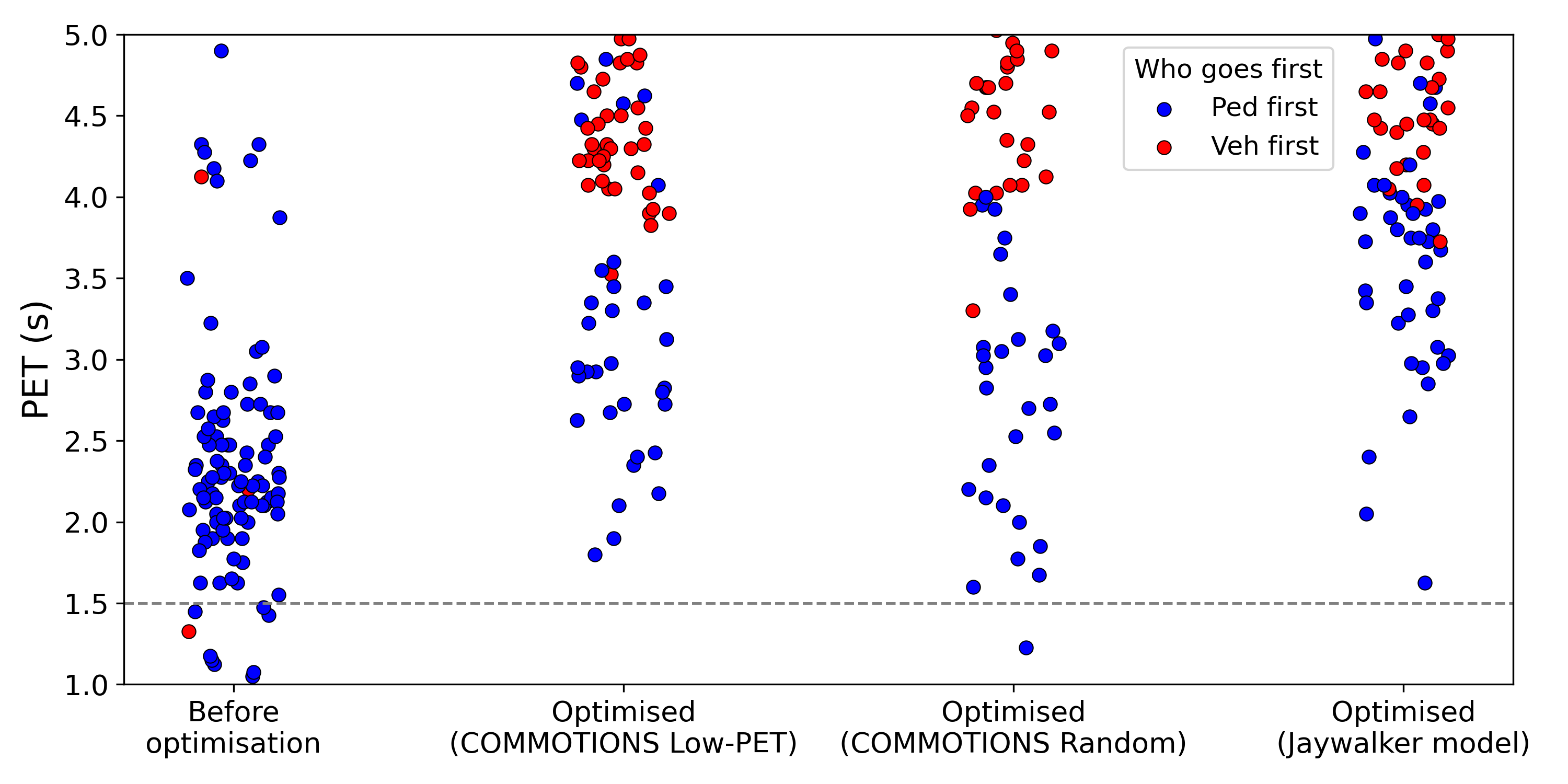}
      \caption{Distribution of PET values across four scenario conditions: Before optimisation, Optimised (Low-PET), Optimised (Random), and Optimised (Jaywalker model). Each point represents an individual scenario, with colours indicating whether the pedestrian or the vehicle passed first. The horizontal dashed line marks the PET safety threshold of 1.5 s. The x axis represents different categories only, and the horizontal spread within each category has no quantitative meaning but is used to improve visual separation of points.}
      \label{fig: comparison_optimised_PET}
\end{figure}

Random-case optimisation also reduced time lost to a similar level (0.04 s) and achieved the lowest sudden speed change rate at 11\%. It further raised mean PET to 5.32 s, higher than the low-PET approach, but its minimum PET (1.23 s) still dropped below the safety threshold, indicating that this optimisation approach did not succeed in eliminating safety-critical interactions. Beyond the summary statistics reported in Table~\ref{tab:av-ped-comparison}, Figure~\ref{fig: comparison_optimised_PET} reveals clear differences in the distributional patterns across conditions. In particular, the random approach produces a denser cluster of PET values close to the 1.5 s threshold, whereas the low-PET approach shifts the entire distribution upward, reducing the near-critical interactions. All three optimisation approaches result in noticeably higher PET values than the before-optimisation baseline.

\begin{table*}[!t]
\centering
\caption{Comparison of AV-pedestrian interaction metrics before and after control parameter optimisation. Optimisation was performed using either low-PET individuals, randomly sampled individuals, or the jaywalker pedestrian model.}
\label{tab:av-ped-comparison}
\small   

\begin{tabularx}{\textwidth}{lYYYY}
\toprule
\textbf{Condition} & \textbf{Mean PET (s)} & \textbf{Min PET (s)} &
\textbf{Sudden speed change rate (\%)} & \textbf{Average vehicle time lost (s)} \\
\midrule
Before optimisation & 2.48 & 1.05 & 83 & 0.12 \\
Optimised\\(COMMOTIONS Low-PET) & 4.96 & 1.80 & 17 & 0.05 \\
Optimised\\(COMMOTIONS Random) & 5.32 & 1.23 & 11 & 0.04 \\
Optimised (Jaywalker model) & 5.52 & 1.62 & 38 & 0.25 \\
\bottomrule
\end{tabularx}
\end{table*}

In contrast to the COMMOTIONS-based optimisation results, optimisation against the jaywalker model raised mean PET to 5.52 s but at a clear cost. Sudden speed changes remained high at 38\%, and average vehicle time lost increased to 0.25 s, more than twice the baseline and far above the results of the COMMOTIONS-based optimisation approaches. Although its minimum PET (1.62 s) indicated that all interactions stayed above the 1.5 s safety threshold, these apparent safety gains arose because the jaywalker model’s exaggerated, highly adversarial behaviour forced the optimiser to select a much larger braking distance. This resulted in an over-cautious AV that braked excessively early, producing artificially high PET values at the expense of comfort and efficiency.

\section{Discussion}
\label{sec:Discussion}

\subsection{Influence of human-like pedestrian model on AV testing}

The results show that using a human-like pedestrian model in AV simulation has a clear impact on the interaction outcomes. Compared with the rule-based CARLA pedestrian, the COMMOTIONS model exhibited gap acceptance behaviour that increased with TTA in line with observed human crossing patterns. We also found that this was accompanied by more human-like vehicle responses: smoother decelerations, fewer abrupt braking events, and PET values that were more consistent across conditions. These findings demonstrate that replacing simplistic scripted agents with human-like agents produces interactions that better reflect real-world dynamics.

This finding has direct implications for simulation-based AV evaluation. The goal of virtual testing is to expose AVs to situations that resemble those encountered in the real world; using unrealistic pedestrian models risks creating interactions that either exaggerate dangers or overlook real risks, leading to results that do not transfer well to deployment. 

\subsection{Implications for AV testing pipelines}

Another contribution of this work is the demonstration that human-like pedestrian models can be used not only to reproduce realistic interactions, but also to generate challenging scenarios that remain behaviourally plausible. The adversarial interactions emerging from the COMMOTIONS model did not require exaggerated or `suicidal' behaviours; rather, they arose naturally from inter- and intra-individual variability in the decision-making process, as well as from the initial spatial–temporal relationship between the pedestrian and the vehicle. This finding suggests that unsafe interactions are not universal but occur with higher probability for certain pedestrian individuals and under certain timing conditions, highlighting the importance of capturing such variability in human behaviour in AV testing. Yet, these naturally occurring interactions were sufficient to expose limitations in AV control logic, highlighting the potential of such models to support stress testing.

The optimisation experiments show that adversarial interactions can also be used to improve AV control performance. Relative to the pre-optimisation controller, braking distance tuning on COMMOTIONS scenarios increased PET, reduced sudden braking, and lowered time lost, demonstrating gains in both safety and efficiency. Compared with optimisation using COMMOTIONS random approach, optimisation using low-PET approach further raised the minimum PET and reduced the number of near safety-critical interactions near the 1.5 s PET threshold, showing that focusing optimisation on safety-critical situations helps the AV avoid rare but high-risk interactions.

A comparison with the jaywalker model further underscores this point. Jaywalker-based optimisation also increased PET but at the expense of comfort and efficiency, as the AV adopted overly defensive action triggered by unrealistic behaviours. In contrast, optimisation using COMMOTIONS adversarial scenarios achieved safety gains without sacrificing comfort or efficiency, yielding smoother decelerations and lower time lost. This demonstrates that human-like adversarial interactions can support controller tuning that balances safety, comfort, and operational efficiency, an outcome directly relevant to AV development pipelines.

By shifting adversarial testing from pure robustness checks towards optimisation, this approach highlights the potential of human-like adversarial scenarios for simulation-based evaluation and controller tuning.

\subsection{Limitations and future work}
Although the present study provides a controlled and systematic evaluation of pedestrian-AV interactions, several limitations remain that point to avenues for future research. First, the experimental setup was restricted to a simplified zebra crossing with a single pedestrian and a single vehicle. While this scenario provides analytical clarity, it does not capture the complexity of urban traffic environments where multiple agents, occlusions, and contextual cues such as road geometry or traffic signals shape decision-making. Extending the framework to multi-pedestrian, multi-vehicle, or shared-space settings would allow a richer assessment of AV performance under realistic conditions.

Second, the AV side of the interaction was limited to two controllers: the CARLA AV controller and the Autoware autonomous driving stack. Although these provide useful baselines, they do not represent the diversity of state-of-the-art AV architectures, such as deep RL-based controllers \citep{lopez2018comparing,kurzer2021generalizing,zhan2023enhance}, or commercial proprietary stacks such as including Tesla Autopilot, Mercedes-Benz Drive Pilot, and the Waymo’s Driver \citep{shadab2025comparison}. Future work would therefore test whether the benefits of behaviourally realistic adversarial scenarios generalise across different control algorithms.

Third, the optimisation targeted only a single parameter—braking distance. Real-world AV safety and comfort depend on a broader set of parameters, including acceleration and jerk limits, gap acceptance thresholds, and the interaction between longitudinal and lateral control. Exploring multi-parameter or adaptive optimisation approaches could yield richer insights into the trade-offs between safety, comfort, and efficiency. Moreover, incorporating more advanced optimisation methods could better support multi-parameter optimisation while reducing computational cost.

Finally, although the COMMOTIONS model captures perceptual, motor, and cognitive variability, it still omits many contextual and social factors that shape pedestrian behaviour, such as group dynamics, cultural norms, or explicit communication signals (e.g. eye contact). It also requires considerable computational resources for parameter sampling and Bayesian optimisation, which may limit scalability to very large test suites. Future work could therefore focus on developing more human-like behavioural models, leveraging parallel simulation infrastructure, and validating against larger naturalistic datasets, in order to strengthen both the robustness and the generalisability of the proposed approach.

\section{Conclusions}
\label{sec:Conclusions}
In this study, we investigated how behaviourally realistic pedestrian models can enhance autonomous vehicle (AV) testing and optimisation. We developed a framework that integrates the COMMOTIONS pedestrian model with adversarial scenario generation and AV controller optimisation in the CARLA simulation environment.

Replacing the rule-based CARLA pedestrian with the COMMOTIONS model yielded more naturalistic interactions, including smoother decelerations, fewer abrupt braking events, and gap acceptance patterns consistent with human data. The adversarial scenario search further identified safety-critical yet plausible interactions arising from inter- and intra-individual variability in pedestrian decision processes and timing, without relying on exaggerated or unrealistic behaviours.

Optimising AV control on these human-like adversarial scenarios improved both safety and efficiency, while avoiding the excessive caution induced by artificially adversarial agents. These results demonstrate that human-like pedestrian models can provide realistic challenges and informative feedback for controller tuning. The proposed framework offers a foundation for scalable and human-centred simulation pipelines, supporting realistic virtual evaluation of AV behaviour across diverse scenarios and control architectures.

\section*{Acknowledgments}
This work was supported by funding from the UK Engineering and Physical Sciences Research Council (EPSRC), grant reference EP/X52573X/1.

\section*{Declarations of interest: none}


\clearpage
\bibliographystyle{elsarticle-harv} 
\bibliography{cas-refs}





\end{document}